\documentclass{ws-ijgmmp}
\usepackage{tensor}
\usepackage{hyperref}
\begin{document}

\markboth{Authors' Names}
{Instructions for Typing Manuscripts (Paper's Title)}

%
\catchline{}{}{}{}{}
%

\title{Inhomogeneous solutions in $f(T,B)$ gravity}

\author{Sebasti\'an N\'ajera}
\address{Instituto de Ciencias Nucleares, Universidad Nacional Aut\'onoma de M\'exico, 
Circuito Exterior C.U., A.P. 70-543, M\'exico D.F. 04510, M\'exico.\,\\
\email{sebastian.najera@correo.nucleares.unam.mx} }

\author{Aram Aguilar}

\address{Instituto de Ciencias Nucleares, Universidad Nacional Aut\'onoma de M\'exico, 
Circuito Exterior C.U., A.P. 70-543, M\'exico D.F. 04510, M\'exico.\,\\
\email{armandoaram$\_$94@hotmail.com} }

\author{Geovanny A. Rave-Franco}

\address{Instituto de Ciencias Nucleares, Universidad Nacional Aut\'onoma de M\'exico, 
Circuito Exterior C.U., A.P. 70-543, M\'exico D.F. 04510, M\'exico.\,\\
\email{geovanny.rave@ciencias.unam.mx} }

\author{Celia Escamilla-Rivera}

\address{Instituto de Ciencias Nucleares, Universidad Nacional Aut\'onoma de M\'exico, 
Circuito Exterior C.U., A.P. 70-543, M\'exico D.F. 04510, M\'exico.\,\\
\email{celia.escamilla@nucleares.unam.mx} }

\author{Roberto A. Sussman}

\address{Instituto de Ciencias Nucleares, Universidad Nacional Aut\'onoma de M\'exico, 
Circuito Exterior C.U., A.P. 70-543, M\'exico D.F. 04510, M\'exico.\,\\
\email{sussman@nucleares.unam.mx} }

\maketitle

\begin{history}
\received{(Day Month Year)}
\revised{(Day Month Year)}
\end{history}

\begin{abstract}
In this paper we explore the possibility to find exact solutions for Teleparallel Gravity (TG) of the type of spherically symmetric Lema\^\i tre-Tolman-Bondi (LTB) dust models. We apply to the LTB metric the formalism of Teleparallel Gravity in its extension to $f(T,B)$ models, which can be seen it as the analagous from the Schwarzschild solution in General Relativity. An exact LTB solution is obtained which is compatible with a specific $f(T,B)$ model whose observational constraints are cosmological viable in a standard spatially flat Robertson-Walker geometry.
\end{abstract}


\section{Introduction}	

Cosmology have been enhanced in the last decade from an enormous increase of new observational surveys whose methods for handling them are still a challenging endeavor, not only because the fundamental nature of the dark sector is still an open question, but also from the persisting \textit{tensions} between the values measured/inferred 
of $H_0$ and $f \sigma_8$ and anomalies in early-time observations \cite{DiValentino:2020vhf, di2021realm}.
 
On the local universe side, there is a wide variety of dark energy proposals to explain the cosmic accelerated  expansion under the General Relativity (GR) framework while considering gauge invariant cosmological perturbations on a Friedmann-Lema\^\i tre-Robertson-Walker (FLRW) background. 
As it is standard, this acceleration behaviour can be
obtained by assuming, for example, a positive cosmological constant $\Lambda$, extra \textit{dark fluids} with negative pressure, scalar fields or modifications of GR, among others \cite{Clifton:2011jh, ishak2019modified}. In this latter line of thought,
extensions of Teleparallel Gravity (TG) \cite{Cai:2015emx, Hohmann:2019nat} have also been considered to explain the cosmic acceleration, and furthermore, to even alleviate the persisting cosmological tensions, all this from the geometry of the theory itself, i.e. without evoking exotic fluids or $\Lambda$ \cite{Briffa:2020qli}. Even more, TG offers several viable cosmological proposals that have show tight constraints using current observations \cite{Nunes:2018xbm,Escamilla-Rivera:2019ulu,Cai:2019bdh}.

TG is a theory of gravity \cite{bahamonde2021teleparallel} in which the curvature-based description of gravity, used in GR, is enhanced by torsion, which involves replacing the Levi-Civita connection with its Weitzenböck connection analog. The name of \textit{teleparallelism} or \textit{distant parallelism} comes from the geometrical description of vectors in the Weitzenböck space-time \cite{Pereira:2013qza}, where torsion can change the direction of such vectors \cite{beltran2019geometrical}. Besides, TG is a gauge theory for the translations group \cite{de2002teleparallel}, so it is possible to represent the gravitational field with a translational gauge potential that appears within the non-trivial part of the tetrad field, i.e. keeps torsion different from zero.

The noteworthy complexity of the Einstein equations and the ones from modified theories of gravity lead us to employ new numerical methods to solve them when evolving with the rest of the matter/energy content of the Universe. Analytical and semi-analytical solutions provide 
useful \textit{toy scenarios} to obtain asymptotic descriptions of cosmic structures using $\Lambda$CDM as background. 

The simplest exact solutions providing this dust-like description are the spherically symmetric Lemaître-Tolman-Bondi (LTB) models  \cite{krasinski1997inhomogeneous,plebanski2006introduction}, whose metric also admits non-trivial solutions with non-zero pressure \cite{sussman2008quasi}. These models were used to explain cosmic acceleration through the effect of large-scale inhomogeneity \cite{enqvist2008lemaitre}. These models were so-called
\textit{Big Void}, since they assumed our location near the center of a giant 1 Gpc spherical void. These solutions have also been very useful as \textit{toy models} to study structure formation \cite{lake2000gravitational, sussman2009quasilocal, deshingkar2001gravitational, bolejko2010structures}, as they can also be conceived as exact non-linear perturbations of a FLRW background, reducing in the linear regime to standard cosmological perturbations in the isochronous comoving gauge \cite{sussman2008quasi,sussman2015spherical}. 

According to the above ideas, in this paper we study the compatibility between TG theories and LTB models. While there is an extensive literature on LTB models in GR,  only a few consider inhomogeneous solutions in TG theories, e.g to analyse the possibility to describe the dark energy dynamics without introducing a cosmological constant \cite{Cai:2015emx} and discuss the problem of energy-momentum localization in TG \cite{salti2005energy}. Our main interest is to examine how the known properties of LTB solutions can be affected by the \textit{torsion} contribution acting as a force in TG theories. Along this work, we refer to quantities with/without a white circle those upon them are calculated with the Levi-Civita/Weitzenböck connection, respectively. The signature used is $(-,+,+,+)$. Also, we will consider geometric units ($c = G = 1$). 

The paper is organized as follows. In Sec.~\ref{sec:LTBGR} we introduce the LTB models constructed from the Schwarzschild solution. In Sec.~\ref{sec:TGR} we described the main principles of TG, the Weitzenb\"ock gauge and a tetrad formalism which obeys this gauge and maintains the spherical symmetry. In Sec.~\ref{sec:LTBTGeq} we discuss LTB-like solutions in TG for a couple of $f(T,B)$ models and their cosmological implications. Finally, in Sec.~\ref{sec:conc} we present concluding remarks.

\section{From Schwarzschild to LTB solutions: the GR approach}
\label{sec:LTBGR}

The Schwarzschild metric in the standard static coordinates can be read as
\begin{equation}
ds^2=-\left(1-\frac{2M_0}{r}\right)dt^2 + \left(1-\frac{2M_0}{Y}\right)^{-1} dY^2 + Y^2d\Omega^2,
\end{equation}
where $M_0$ denotes the Schwarzchild mass. The radial timelike geodesics of this spacetime follows
\begin{equation} \dot{Y}^2 = \frac{2M_0}{Y}-\kappa,\quad \dot t = \frac{\sqrt{1-\kappa}}{1-2M_0/Y}, \quad \kappa \equiv \frac{\kappa_0^2}{4}-1,\label{rtdots}\end{equation}
where $\kappa_0$ is the conserved energy in each geodesic. A new coordinate system $(\tau,r)$ based on the world lines of observers following radial geodesics can be constructed from Eq.~(\ref{rtdots}) \cite{plebanski2006introduction}. Under this coordinate system the proper time $\tau$ is the time coordinate, while $r$ is a continuous variable that label each geodesic. As the binding energy $k$ is a continuous parameter that varies along the geodesic congruence in the $(\tau,r)$ plane, it can be shown that $\kappa=\kappa(r)$ and $Y=Y(\tau,r),\,\,t=t(\tau,r)$. The coordinate transformation $(t,Y)\to (\tau,r)$ obtained by Eq.~(\ref{rtdots}) takes the Schwarzschild static metric into the following time dependent form:
\begin{equation} ds^2 = -d\tau ^2+\frac{Y'^2}{1-\kappa}\,dr^2+Y^2\,d\Omega^2,\label{ScwLTB}\end{equation}
where $\kappa=\kappa(r)$ and $Y'\equiv \left[\partial Y/\partial r\right]_\tau$, with $Y(\tau,r)$ satisfying
\begin{eqnarray}
&&\dot{Y}^2 =\frac{2M_0}{Y}+\kappa,\label{rdot}\\ &&
\quad \tau-\tau_B(r)=\int_{\bar{Y}=0}^Y{\frac{\sqrt{\bar{Y}}d\bar{Y}}{\sqrt{2M_0+\kappa(r)\,\bar{Y}}}},\nonumber
\end{eqnarray}
where $\dot{Y}\equiv u^\mu Y_{,\mu}= \left[\partial Y/\partial \tau\right]_r$ \cite{ellis2012relativistic} and $\tau_B(r)$ an integration constant characterizing the value of $\tau$ for which $Y=0$, i.e. the equivalent to the Big Bang singularity in FLRW models. In these coordinates notice that the Schwarzschild radius $Y_s=2M_0$ is a regular point.

The Schwarzschild mass, $M$, can be interpreted in spherical coordinates of Eq.~\eqref{ScwLTB} as an integrated distribution $M_0 =\int_V{\rho\,d^3 x}$ where $\rho=M_0\delta^3(0)/(r^2\sin\theta)$. By considering a continuous density instead of a distributional density (which is equivalent to considering $M_0 \, \rightarrow \, M=M(r)$) and taking as the $4$-velocity $u^\mu =\left[dx^\mu/d\tau\right]_{r}= \delta^\mu_\tau$, for observers comoving along the geodesics. 
The substitution of Eqs.~\eqref{ScwLTB}--\eqref{rdot} in the Einstein equations, $G_{\mu\nu}=8\pi \Theta_{\mu\nu}$ yields
\begin{equation} \Theta^{\mu\nu} =\rho\,u^\mu\,u^\nu=\rho\,\delta^\mu_\tau\,\delta^\nu_\tau, \quad  8\pi\rho = \frac{2M'}{Y^2\,Y'},\quad u^\mu=\delta^\mu_\tau. \label{eq:charLTB}
\end{equation}
These relations characterize LTB models with a dust source. 
The LTB metric has a natural FLRW background, defined in the space of parameters by the election of the free functions in the following way \cite{krasinski1997inhomogeneous}
\begin{equation}\label{eq:FLRWGR}
    Y=ra(\tau),\quad \kappa=kr^2, \quad M=\frac{H_0^2\Omega_{M0}r^3}{2},
\end{equation}
where $a(\tau)$ is the FLRW scale factor in conformal time, $\Omega_{m0}=\Omega_{m}(\tau)_{\tau=\tau_0}$ is the matter density parameter evaluated at current time and $\tau_0$ is current time defined by $a(\tau_0)=1$. $H_0=H_{\tau=\tau_0}=\dot a(\tau)/a(\tau)_{\tau=\tau_0}$ is the Hubble parameter and $k$ is the constant curvature of the space--like hypersurfaces orthogonal to $u^a$. In this way, Eq.~\eqref{rdot} in the FLRW limit Eq.~\eqref{eq:FLRWGR} reads
\begin{equation}\label{eq:Fried}
    \frac{\dot a^2}{H_0^2a^2}=\frac{H^2}{H_0^2}=\frac{\Omega_{M0}}{a^3}+\frac{\Omega_{k,0}}{a^2},
\end{equation}
with $\Omega_{k,0}$ the spatial curvature density.


\section{Teleparallel Gravity on basis}
\label{sec:TGR}

TG arises as a gauge theory of the group of translations based on Noether's Theorem since the energy–momentum current is covariantly conserved if the Lagrangian is invariant under space-time translations \cite{Krssak:2018ywd,bahamonde2021teleparallel}, and is a geometric theory of gravity where the gravitational field is mediated by the torsion and not by the curvature, so it requires that the components of the $2$-form curvature vanish identically. The dynamical field in this case is the tetrad, and is related to the gauge potential of the translations group \cite{Golovnev:2017dox}. The tetrad and the metric are related by the following expression
\begin{align}
g_{\mu \nu} = \eta_{AB}e^{A}_{\enskip \mu}e^{B}_{\enskip \nu}. \label{eq:MetricAndTetrad}
\end{align}
As we can notice, if we perform a local Lorentz transformation on the tetrad $e^{A}_{\enskip\mu} \to e^{B}_{\enskip\mu}\Lambda(\mathbf{x})_{B}^{\, A}$, the metric remains invariant under such transformations. 

Since different tetrads related upon local Lorentz transformation give the same metric, TG has to be local Lorentz invariant \cite{Aldrovandi2013}. This requirement
gives as a consequence an extra purely inertial degree of freedom called the \textit{spin connection} $\omega_{\enspace B\mu}^{A}$, which is a $1$-form that assumes values in the Lie Algebra of the Lorentz group \cite{Aldrovandi2013}. In this context, TG consist of a geometrical setup that includes a manifold $M$, a tetrad $\mathbf{e} 
= \{e^A\}^{3}_{A=0}$ and a spin connection $\boldsymbol{\omega}=\{\omega^A_{\hspace{0.15cm}B}\}_{A,B=0}^3$ \cite{Hohmann:2019nat}. 
Once a coordinate chart $\{x^\mu\}$ is given on the spacetime manifold $M$ and a second one $\{x^A\}$ is given on the Minkowski space as a fiber bundle on $M$, we can write 
the tetrad and the spin connection, respectively as
$e^A = e^{A}_{\enspace \mu}dx^{\mu}$ and $\omega^A_{\hspace{0.15cm}B} = \omega^A_{\hspace{0.15cm}B\mu}dx^{\mu}$ \cite{Aldrovandi2013,Hohmann:2019nat}.
In TG, the linear connection on $M$ is no longer the Levi-Civita connection but the Weitzenböck connection defined by
\begin{align}
\Gamma_{\enspace \nu \mu}^{\sigma} := E_{A}^{\enspace \sigma}\partial_{\mu}e_{\enspace \nu}^{A} + E_{A}^{\enspace \sigma}\omega_{\enspace B\mu}^{A}e_{\enspace \nu}^{B},\label{eq:WeitzenbockConnection}
\end{align}
where $e_{\enspace \mu}^{A}$ is the tetrad field, $E_{A}^{\enspace \sigma}$ the transpose tetrad field defined by $E_{A}^{\enspace \mu}e_{\enspace \nu}^{A}=\delta^{\mu}_{\nu}$. Greek indices refer to space-time coordinates while capital Latin letters stand for tangent space indices. The condition that the components of the $2$-form curvature vanish identically can only be achieved by the purely inertial spin connection \cite{Krssak:2015oua, Hohmann:2019nat}
\begin{align}
\omega_{\enspace B\mu}^{A} = \Lambda(\mathbf{x})^{A}_{\enspace C}\partial_{\mu}\Lambda(\mathbf{x})^{C}_{\enspace B}. \label{eq:PureIneretialSpin}
\end{align}
The spin connection given by Eq.~(\ref{eq:PureIneretialSpin}) can be chosen to satisfy the Weitzenböck gauge $\omega_{\enspace B\mu}^{A} =0$ if a local Lorentz transformation is performed, since it is a transformed zero spin connection in an arbitrary Lorentz frame \cite{Pereira:book} and then, the tetrad becomes the only fundamental field. This approach is known as the \textit{pure tetrad formalism} \cite{Krssak:2018ywd,bahamonde2021teleparallel}, which breaks the local Lorentz invariance of the theory and can led us to a sensible choice of the tetrad field on its extensions. This issue is so-called  \textit{the choice of good and bad tetrads} \cite{Golovnev:2021omn,Li:2010cg,Tamanini:2012hg}.

\subsection{The Teleparallel Equivalent theory}

The action in TG theories has a simple case that is equivalent to GR, known as the Teleparallel Equivalent to General Relativity (TEGR), which Lagrangian density takes the form $-T+B$, and the variation  of the action  with  respect  to  the  tetrad  field  results completely equivalent to the GR dynamical equations \cite{Bahamonde:2019jkf}.
Raising the TEGR action to its $f(T,B)$ gravity extension results in
\begin{equation}\label{fTBaction}
\mathcal{S} = \int d^4x
\bigg[\dfrac{1}{16\pi}f(T,B) + \mathcal{L}_{\text{ m}}\bigg]e,
\end{equation}
where $e$ is the determinant of the tetrad field $e = \det (e^{A}_{\enskip \mu}) = \sqrt{-g}$, $\mathcal{L}_{\text{ m}}$ represents the Lagrangian for matter and $f=f(T,B)$ is an arbitrary function of $T$ and $B$. $T$ is the torsion scalar and $B$ is the boundary term defined by
\begin{equation}
T := S_{\alpha}^{\enspace \sigma \rho}T_{\enspace \sigma \rho}^{\alpha}, \quad B := \frac{2}{e}\partial_{\mu}\Big(e\tensor{T}{^\nu_\nu^\mu}\Big),
\end{equation}
where the superpotential and the torsion tensor are given by
\begin{equation}\label{superpotential}
S_{\alpha}^{\text{   }\sigma \rho} = \frac{1}{2}(T_{\alpha}^{\text{   }\sigma \rho} + T^{\rho\sigma}_{\hspace{0.3cm} \alpha} - T^{\sigma \rho}_{\hspace*{0.3cm}\alpha} - 2 T^{\lambda\sigma}_{\hspace*{0.3cm}\lambda}\delta_{\alpha}^{\rho} + 2 T^{\lambda \rho}_{\hspace*{0.3cm}\lambda}\delta_{\alpha}^{\sigma}),
\end{equation}
\begin{equation}\label{torsiontensor}
T_{\enspace \sigma \rho}^{\alpha}:=  2\Gamma_{\enspace [\rho \sigma]}^{\alpha}.
\end{equation}

This generalization of TG provides a richer class of models and may recover $f(\mathring{R})$ models if the functional dependence takes the form $f(T,B)= f(-T + B)$.
Taking a variation of Eq.~(\ref{fTBaction}) with respect to the tetrad field leads to \cite{bahamonde2021teleparallel}

{\normalsize
\begin{align}
    E_{A}^{\enskip\mu}\mathring{\Box}f_{B}-E_{A}^{\enskip\nu}\mathring{\nabla}^{\mu}\mathring{\nabla}_{\nu}f_{B}+\frac{1}{2}Bf_{B}E_{A}^{\enskip\mu}-\partial_{\nu}(f_{B}+f_{T})S_{A}^{\enskip\mu\nu}-\frac{1}{e}f_{T}\partial_{\nu}(eS_{A}^{\enskip\mu\nu})\nonumber\\
    +f_{T}T^{B}_{\enskip\nu A}S_{B}^{\enskip\nu\mu}-f_{T}\omega^{B}_{\enskip A\nu}S_{B}^{\enskip\nu\mu}-\frac{1}{2}fE_{A}^{\enskip\mu}=8\pi G\Theta_{A}^{\enskip\mu},\label{fieldFTB}
\end{align}
}
where $\mathring{\Box} = \mathring{\nabla}^{\delta}\mathring{\nabla}_{\delta}$, $\Theta_{A}^{\enskip\mu}$, is the energy-momentum tensor and $f_{T}$ and $f_{B}$ refer to the derivative of $f$ with respect to the torsion scalar and boundary term, respectively. 

The variation of the action Eq.~\eqref{fTBaction} with respect to the spin connection leads to the requirement that the antisymmetric part of the field equations must vanish. 
Hence, a good pair of tetrad-spin connection satisfies the symmetric part of the field equations Eq.~\eqref{fieldFTB}. In addition, the antisymmetric part must vanishes \cite{bahamonde2021teleparallel,Bahamonde:2021srr}.

\subsection{The tetrad-spin connection pair}

When we work with extensions of TEGR, the symmetric and antisymmetric part of the field equations Eq. \eqref{fieldFTB} must be solved simultaneously to find a good pair of tetrad-spin connection. Setting a general recipe 
to obtain a spin connection in terms of the tetrad is non-viable, since both are independent fields. However, in \cite{Hohmann:2019nat} was found that for some symmetries it is possible to find a pair of tetrad-spin connection which makes the torsion tensor exhibit the required symmetry, even in the Weitzenböck gauge. 
On this path, we will set up these ideas into a spherical symmetric case. Based on the study of space time symmetries from Cartan geometry, a symmetry can be seen as a group action $\varphi: G \times M \to M$ of a Lie Group $G$ such that $\varphi^*_u g = g$ and $\varphi^*_u \Gamma = \Gamma$ for all $u \in G$, where $g$ is the induced metric from the tetrad \eqref{eq:MetricAndTetrad} and $\Gamma$ the induced connection from the tetrad and the spin connection \eqref{eq:WeitzenbockConnection}.

Hence, a teleparallel geometry is invariant under $\varphi$ if there exists a local Lie group homomorphism $\mathbf{\Lambda}:G \times M \to \text{SO}(1,3)$ such that certain conditions are satisfied. When working with infinitesimal symmetries, those conditions are translated in terms of Lie derivatives as
\begin{align}
(\mathcal{L}_{X_{\xi}}e)^A_{\;\mu} = - \boldsymbol{\lambda}^A_{\;\xi B}e^B_{\mu}, \quad (\mathcal{L}_{X_{\xi}}\omega)^A_{\; B \mu} = \text{D}_{\mu}\boldsymbol{\lambda}^A_{\;\xi B}, \label{eq:symmetryconditions}
 \end{align}
 where $\text{D}_{\mu}\boldsymbol{\lambda}^A_{\;\xi B} = \partial_{\mu}\boldsymbol{\lambda}^A_{\;\xi B} + \omega^{A}_{\; C \mu}\boldsymbol{\lambda}^C_{\;\xi B} - \omega^{C}_{\; B \mu}\boldsymbol{\lambda}^A_{\;\xi C}$, $\boldsymbol{\lambda}$ is the local Lie algebra homomorphism defined in terms of derivatives of $\mathbf{\Lambda}$. By performing a local Lorentz transformation to the Weitzenböck gauge, the symmetry conditions Eq.~\eqref{eq:symmetryconditions} read as
 \begin{align}
(\mathcal{L}_{X_{\xi}}e')^A_{\;\mu} = - \boldsymbol{\lambda}'^A_{\;\xi B}e^B_{\mu}, \quad 0 \equiv (\mathcal{L}_{X_{\xi}}\omega')^A_{\; B \mu} = \partial_{\mu}\boldsymbol{\lambda}'^A_{\;\xi B}.
 \end{align}
Therefore, the local Lie group homomorphism $\mathbf{\Lambda}'_u$ and the local Lie algebra homomorphism $\boldsymbol{\lambda}'_{\xi}$ do not depend on the space time point, so the Lie group homomorphism is not local but global. Then, in order to find the tetrad which is compatible with the Weitzenböck gauge, we only have to choose a global homomorphism and solve the symmetry condition Eq.~\eqref{eq:symmetryconditions} for the tetrad. By doing so with a spherical symmetric space time, it is found that the most general $\text{SO}(3)$ tetrad compatible with the Weitzenböck gauge is
{\footnotesize\begin{align}
(e^A{}_{\mu}) = \begin{pmatrix}
C_1 & C_2 & 0 & 0 \\
C_3 \sin \theta \cos \phi & C_4 \sin \theta \cos \phi & C_5\cos \theta \cos \phi - C_6 \sin \phi & -\sin \theta (C_5 \sin \phi + C_6 \cos \theta \cos \phi) \\
C_3 \sin \theta \sin \phi & C_4 \sin \theta \sin \phi & C_5\cos \theta \sin \phi + C_6 \cos \phi & \sin \theta (C_5 \cos \phi - C_6 \cos \theta \sin \phi) \\ 
C_3 \cos \theta & C_4 \cos \theta & -C_5\sin \theta & C_6 \sin^2\theta
\end{pmatrix},
\end{align}}
where the induced metric has non-vanishing components $ g_{tt}= C_3^2  -  C_1^2$ , $g_{rr} = C_4^2 - C_2^2$, $g_{tr}=g_{rt} = C_3C_4 - C_1C_2$ , $g_{\theta \theta} = C_5^2 + C_6^2$ and $g_{\phi \phi} = g_{\theta \theta}\sin^2 \theta$. In the standard spherical symmetric spacetime, only the $tr$ and $\theta \phi$ components of the antisymmetric part of the field equations, using this tetrad, are non-vanishing and have to be solved simultaneously with the symmetric part of the field equations \cite{Hohmann:2019nat,bahamonde2021teleparallel}. By choosing $C_1 = 1$, $C_2 = C_3 = C_6= 0$ , $C_4 = \frac{Y_r}{\sqrt{1 - \kappa}}$ and $C_5 = Y$, we recover the LTB metric Eq.~\eqref{ScwLTB} and then, the LTB tetrad compatible with the Weitzenböck gauge is
\begin{align}\label{tetrad-LTB-Wgauge}
(e^A{}_{\mu})_{ \text{LTB}} = \begin{pmatrix}
1 & 0 & 0 & 0 \\
0 & \frac{Y_r}{\sqrt{1 - \kappa}} \sin \theta \cos \phi & Y\cos \theta \cos \phi  & -Y\sin \theta \sin \phi  \\
0 & \frac{Y_r}{\sqrt{1 - \kappa}} \sin \theta \sin \phi & Y\cos \theta \sin \phi & Y\sin \theta \cos \phi  \\ 
0 & \frac{Y_r}{\sqrt{1 - \kappa}} \cos \theta & -Y\sin \theta & 0
\end{pmatrix}.
\end{align}
This is the tetrad field we are going to consider in our study.
 Although we will be working on the Weitzenböck gauge, it is worth saying that it is possible to work with a diagonal tetrad and a non-vanishing spin connection. Such scenario can be achieved by performing a local Lorentz transformation on the tetrad and also on the spin connection through the Eq.~\eqref{eq:PureIneretialSpin}, with $\Lambda^A_{\; \; C}(\mathbf{x})$ the same local Lorentz transformation used on the tetrad to transform it into a diagonal tetrad. The advantage of working without the Weitzenböck gauge is that we can deal with a diagonal tetrad, but  we will have a spin connection with components different from zero, as a consequence. Since the field equations are invariant under local Lorentz Transformations, it is a matter of choice to work with or without the Weitzenböck gauge.


\section{From Schwarzschild to LTB: The Teleparallel approach}
\label{sec:LTBTGeq}

With the choice of tetrad Eq. \eqref{tetrad-LTB-Wgauge} and considering a dust source $\Theta_{\mu\nu}=\rho u_\mu u_\nu$, with $\rho$ the matter--energy density, the diagonal components of Eq.~(\ref{fieldFTB}) can be read as:

{\tiny{\begin{eqnarray}
f-Bf_B+\left(\frac{\kappa'}{Y'^2}+\frac{2(1-\kappa)Y''}{Y'^3}-\frac{4\sqrt{1-\kappa}}{Y'Y}\right)f_B'+\left(\frac{2\dot{Y}'}{Y'}+\frac{4\dot{Y}}{Y}\right)\dot f_B+4\frac{1-\kappa-\sqrt{1-\kappa}}{Y'Y}f_T'&\nonumber\\
+2\left(\frac{2(1-\kappa-\sqrt{1-\kappa}-\dot{Y}^2)}{Y^2}-\frac{4\dot{Y}'\dot{Y}+\kappa'}{Y'Y}
\right)f_T+\frac{2(-1+\kappa)}{Y'^{2}}f_B''=16\pi\rho,\quad &\label{eq:eqtt}\\
\left(\frac{\ddot Y Y'^2+\dot Y'\dot YY'}{(1-\kappa)Y}+\frac{(\dot Y^2-(1-\kappa))\sqrt{1-\kappa}+1-\kappa))Y'^2}{(1-\kappa)^{\frac32} Y^2}\right)f_T+\frac{Y'}{Y}f_B'+\frac{\dot YY'^2}{(1-\kappa)Y}\dot f_T+\frac{(Bf_B-f-2\ddot f_B)Y'^2}{4(1-\kappa)}=0,\quad &\label{eq:eqrr}\\
\left(\frac{\sqrt{1-\kappa}Y}{Y'}-\left(\frac{Y''(1-\kappa)}{Y'^3}+\frac{\kappa'}{2Y'^2}\right)Y^2\right)f_B'+\left(\frac{\ddot Y'Y^2}{Y'}+\left(\ddot Y+\frac{6\dot Y'\dot Y+\kappa'}{2Y'}\right)Y+\dot Y^2+2\sqrt{1-\kappa}+\kappa-2\right)f_T\nonumber&\\
+\frac{(\sqrt{1-\kappa}-(1-\kappa)Y}{Y'}f_T'+\left(\frac{\dot Y'Y^2}{Y'}+\dot Y Y\right)\dot f_T+\left(\frac{1-\kappa}{Y'^2}f_B''-\ddot f_B+\frac12 (Bf_B-f)\right)Y^2=0.\quad &\label{eq:eqthth}
\end{eqnarray}}}
The only non-zero component of the antisymmetric part of the field equations is given by
\begin{eqnarray}\label{eq:antisymgen}
\frac{((\sqrt{1-\kappa}-1)Y'(\dot f_B+\dot f_T)-\dot Y\sqrt{1-\kappa}(f_B'+f_T')}{Y\sqrt{1-\kappa}}=0.\label{eq:antisym}
\end{eqnarray}
From this latter notice that $\dot f_T=-\dot f_B$ and  $f_T'=-f_B'$ are solutions, which imply that $f(T,B)=f(T-B)-T+B=f(R)+\textrm{TEGR}$. 
In order to have a non-trivial form of the functional we must solve Eq.~\eqref{eq:antisym} with a generic choice of $f(T,B)$, which will be the initial point for our following analysis. 

\subsection{An exact LTB solution in $f(T,B)$ gravity: the generic procedure}

To obtain a closed analytic solution we consider the spatially flat subcase $\kappa = 0$, $k$ is proportional to the 3-dimensional Ricci scalar of constant $\tau$ slices. According to this, Eq.~\eqref{eq:antisym} can be read as
\begin{equation}\label{eq:antisymk0}
	\frac{\dot Y(f_B'+f_T')}{Y}=0,
\end{equation}
and the field equations Eqs.~\eqref{eq:eqtt}-\eqref{eq:eqthth} take the following form 
{\footnotesize{\begin{eqnarray}
\left(\frac{Y''}{Y'^3}-\frac{2}{Y'Y}\right)f_B'+\left(\frac{\dot Y'}{Y'}+\frac{2\dot Y}{Y}\right)\dot{f}_T-\left(\frac{4\dot Y\dot Y'Y''}{Y'Y}+\frac{2\dot Y^2}{Y^2}\right)f_T-\frac12\left(Bf_B-f\right)-\frac{f_B''}{Y'^2}&=8\pi \rho,\nonumber\\
\label{eq:eqttk0}\\
\frac{2Y'}{Y}f_B'+\frac{2\dot YY'^2}{Y}\dot f_T+2\left(\frac{\ddot YY'^2+\dot Y\dot Y'Y'}{Y}+\frac{\dot Y^2Y'^2}{Y^2}\right)f_T+\left(\frac12\left(Bf_B-f\right)-\ddot f_B\right)Y'^2&=0,\nonumber\\
\label{eq:eqrrk0}\\
\left(-\frac{Y''}{Y'^3}+\frac{1}{Y'Y}\right)f_B'+\left(\frac{\ddot Y'}{Y'}+\left(\frac{\ddot Y}{Y^2}+\frac{3\dot Y\dot Y'}{Y'Y^2}\right)+\frac{\dot Y^2}{Y^2}\right)\dot{f}_T+\left(\frac{\dot Y'}{Y'}+\frac{\dot Y}{Y}\right)\dot f_T&\nonumber\\+\left(\frac12\left(Bf_B-f\right)-\ddot f_B+\frac{f_B''}{Y'^2}\right)&=0.\nonumber\\
\label{eq:eqththk0}
\end{eqnarray}}}

Solving Eqs.~\eqref{eq:antisymk0}-\eqref{eq:eqttk0} are restricted to the election of the functional $f$ and its associated parameters, without further restricting the functional form of the radial function $Y$.

The selection of the $f$ functional must be based on the physical problem in consideration and observational data. Two physical scenarios we can consider with the LTB geometry are local structure formation and the accelerated expansion of the Universe \cite{enqvist2008lemaitre,bolejko2010structures} in particular dealing with the $H_0$ tension, since we are considering a late time universe scenario according with the Hubble flow. The LTB metric in GR has been widely used to cover these phenomena as it is an exact perturbation of the FLRW metric \cite{sussman2009quasilocal}, therefore several toy models have been constructed to model structure formation and cosmic acceleration \cite{enqvist2008lemaitre,bolejko2010structures}. Local structure formation is based on matter perturbations and gravitational collapse. Usually, the effect of dark energy is neglected in consequence we will consider cosmic acceleration separately. In the following subsection we restrict our study to specific functionals, showing a generic recipe for solving Eqs.~\eqref{eq:eqtt}-\eqref{eq:antisym}, the election of such functional is based on physical assumptions.

We start by describing the general prescription to be followed with any $f(T,B)$ functional for the LTB geometry.  When a dust source is considered, all pressures are zero, although this procedure can easily be extended to s perfect fluid source. Consequently,  Eqs.~\eqref{eq:eqrrk0}-\eqref{eq:eqththk0} must be equal and therefore substracting them will impose additional constraints on the form of the $f$ functional as a partial  differential equation (PDE). Using a chain rule we can write all temporal and radial derivatives of the $f$ functional in derivatives w.r.t $B$ or $T$, that will give two separate PDE's for $Y$, or a single PDE considering $T$ and $B$ depend on $Y$. Solving any of this PDE's is cumbersome, nevertheless we introduce a specific ansatz based on the GR solution and considering that, based on the discussion of Sec.~\ref{ScwLTB}, the Schwarzschild solution is also a solution to the $f(T,B)$ field equations and we will consider an ansatz based the LTB geometry. 

Any $f$ functional of class $\mathcal C^\infty(\mathcal M)$ can be expanded in a power law series
\begin{equation}
	f(T,B)=\sum_i\left(\alpha_i T^i+\beta_iB^i+\sum_j\gamma_{ij}T^iB^j
\right),
\end{equation}
with $\alpha_i,\beta_i, \gamma_{ij}$, constants. Therefore we can consider power law $f(T,B)$ functionals as approximations to any general one. To obtain a natural TEGR limit it is useful to consider $\alpha_1=-1$, as setting all the other constants  $\alpha_i,\beta_i, \gamma_{ij}$ gives the TEGR Lagrangian, and we can always omit the linear term for the boundary term since it does not contribute to the dynamics. Since we would like to obtain a solution that could model structure formation, as well as explaining the cosmic late time acceleration, we use $f(T,B)$ functionals that have proven to be successful to yield late cosmic acceleration without evoking a cosmological constant \cite{Escamilla-Rivera:2019ulu}. Therefore  we consider: a power law functional and a mixed power law functional.

\subsection{Structure formation}
\label{sec:struform}

LTB metrics have been widely used as toy models to explain the structure formation \cite{bolejko2010structures, krasinski2001structure, alonso2010large}. Therefore, we are interested in analyse  the compatibility of LTB geometry within TG. In particular, we will study the possible LTB solutions that denotes structure formation landscapes where the effects of the accelerated cosmic flow are negligible.


\subsubsection{Power Law}

The $f(T,B)$ power law functional is given by
\begin{equation}
\label{PLM}
f(T,B)=-T+\alpha T^m+\beta  B^n,
\end{equation}
with $\alpha, \beta , n, m $ constants to be determined with initial or boundary conditions on the dynamical field equations. Notice that Eq.~\eqref{fTBaction} 
allow us to reduce to the TEGR case when $n=1, \alpha=0, \beta =1$. However, given the previous values but allowing $\beta $ to be a free parameter, the field equations are not affected since we are still having linear boundary terms. Furthermore, it is standard to consider that $n\neq 1$ to have modifications on gravity. To analyse this case, we have to solve the field equations Eqs.~\eqref{fieldFTB} which are the antisymmetric part of the field equations Eq.~\eqref{eq:antisymk0} as well as Eqs.~\eqref{eq:eqttk0}-\eqref{eq:eqththk0}. Using a chain rule we can transform the radial and time derivatives of $f_T$ and $f_B$ in Eq.~\eqref{eq:antisymk0} into a PDE to be solved
{\small
\begin{align}
m(m-1)\beta B^{m-2}\left(\frac12\frac{\partial}{\partial r}\left(\frac{\ddot Y'}{Y'}\right)Y^3+2\frac{\partial}{\partial r}\left(\frac{\dot Y'\dot Y}{Y'}\right)Y^2+\frac{\partial}{\partial r}\left(\ddot Y Y\right)Y -2\frac{\partial}{\partial r}\left(\dot Y Y\right)Y'\right)\nonumber\\+n(n-1)\alpha T^{m-2}\left(\frac{\partial}{\partial r}\left(\frac{\dot Y' \dot Y}{Y'}\right)Y^2-\dot Y^2 Y'\right)&=0.\label{eq:tobesolved}
\end{align}}

Notice that the choices $n=1$ with  $m=0$ or $m=1$ yield an exact solution as these are the TEGR case. As previously described we wish to have a non-linear boundary term, therefore we have to solve the PDE for Y in the parenthesis. Following our discussion from Sec.~\ref{sec:LTBGR}, since the Schwarzschild space--time is a solution in TG,  we introduce the following ansatz
\begin{equation}
\label{eq:Ydotansatz}
	\dot Y^2 = \frac{2M}{Y^l},
\end{equation}
where $M=M(r)$ and $l$ is a constant to be determined. Eq.~\eqref{eq:Ydotansatz} does not yield an accelerated expansion behaviour unless $l<0$, even if the solutions do not admit negative values for the $l$ constant we can always introduce an ansatz that will admit accelerated expansion, we will treat this case in Sec.~\ref{sec:accexp}.

With this ansatz Eq.~\eqref{eq:Ydotansatz} the torsion scalar and the boundary term read, respectively
\begin{eqnarray}
T&=&-\frac{4(M(l-1)Y'-M'Y)}{Y'Y^{l+2}},\\
B&=&-\frac12 (l-4)T.	
\end{eqnarray}
To obtain a behaviour due to a boundary term we need $l\neq 4$. A possible solution is to consider $l=4$, which reduce the equations to a $f(T)$ gravity. In this case, Eq.~\eqref{eq:Ydotansatz} yields a closed form of the radial function $Y$, 
 \begin{equation}\label{eq:exsolYPWFT}
 	Y=\left(3\sqrt{2M}(\tau-\tau_B(r))\right)^{\frac{1}{3}},
 \end{equation}
where $\tau_B$ is the integration constant described in Sec.~\ref{sec:LTBGR}, and this form of Y readily solves Eq.~\eqref{eq:antisymk0}.  
For each choice of the free parameters $n,m,l$ solving Eq.~\eqref{eq:antisymk0} without considering Eqs.~\eqref{eq:eqrrk0}-\eqref{eq:eqththk0} does not imply a consistent exact solution.  
 
For our selected ansatz and the $l=4$ case, both Eqs.\eqref{eq:eqrrk0}-\eqref{eq:eqththk0} are equal. This procedure is also valid for a perfect fluid, as the RHS of Eqs.~\eqref{eq:eqrrk0}-\eqref{eq:eqththk0} will be proportional to the isotropic pressure. Finally $l=4$  solves Eq.~\eqref{eq:antisymk0} without imposing additional constraints in the variables $Y,M$. With these considerations Eq.~\eqref{eq:eqttk0} reads
{\small
\begin{align}
	8\pi \rho=\frac{1}{(M'Y-3MY')^3}\left(\frac{1}{Y'}\left(\frac{162Y'^4M^4}{Y^6}+\frac{2M'^4}{Y^2}-\frac{4^n\alpha\left(n-\frac12\right)(M'Y-3MY')^{3+n}}{Y^{6n}Y'^{n-1}}\right)\right.\nonumber\\
	\left.-\frac{216M'Y'^2M^3}{Y^5}+\frac{108M'^2Y'M^2}{Y^4}-\frac{24M'^3M}{Y^3}\right),
\end{align}}
which will set additional constraints to have positive mass--energy density for every $n$, for example in the particular cases $n=1,2,3$ we have the following form for $\rho$:
\begin{eqnarray}
8\pi\rho = \frac{2M'}{Y'Y^5}-\frac{6M}{Y^6},\\
8\pi\rho = \frac{2M'}{Y'Y^5}-\frac{6M}{Y^6}-\frac{24\alpha M'^2}{Y'^2Y^{10}}+\frac{144\alpha MM'}{Y'Y^{11}}-\frac{216\alpha M^2}{Y^{12}},\\
8\pi\rho = \frac{2M'}{Y'Y^5}-\frac{6M}{Y^6}-\frac{160\alpha M'^3}{Y'^3Y^{15}}+\mathcal O\left(\frac{1}{Y^{16}}\right).
\end{eqnarray}
Notice that the closed form of Eq.~\eqref{eq:exsolYPWFT} allows for a continuous and regular form of the mass--energy density as all $r$ values where $Y=0$ are continuous. We obtain the existence of one center of symmetry (marked by r=0), a fixed point of the SO(3) group. LTB models with no symmetry centers, and two centers of symmetry exist in GR \cite{humphreys2012regular}. The existence of none or two symmetry centers might change some of our results, as has been reported in LTB solutions in f(R) gravity \cite{sussman2017lemaitre}. LTB solutions with these characteristics must be further analyzed and this will be reported elsewhere. Nevertheless it is important to remark that the value $l=4$ must necessarily impact on structure formation, since the ansatz Eq.~\eqref{eq:Ydotansatz} yields the Friedmann equation in a suitable limit as shown in Sec.~\ref{ScwLTB}.

We now consider the particular case in which we do not consider $l=4$ to have a boundary term. To solve Eq.~\eqref{eq:tobesolved} we should consider particular choices of the free parameters of the $f$ functional. We choose the particular case $\alpha=0$, as in \cite{Escamilla-Rivera:2019ulu} indicate that considering at least a quadratic boundary term  will reproduce several cosmological scenarios such as cosmic acceleration. Consequently, we can solve 
\begin{equation}
	\left(\frac12\frac{\partial}{\partial r}\left(\frac{\ddot Y'}{Y'}\right)Y^3+2\frac{\partial}{\partial r}\left(\frac{\dot Y'\dot Y}{Y'}\right)Y^2+\frac{\partial}{\partial r}\left(\ddot Y Y\right)Y -2\frac{\partial}{\partial r}\left(\dot Y Y\right)Y'\right)=0.\label{eq:tbsBPW}
\end{equation}
As stated above, we introduce the ansatz Eq.~\eqref{eq:Ydotansatz}, which gives as only solution $l=1$. Moreover, the subtraction of Eqs.~\eqref{eq:eqrrk0}-\eqref{eq:eqththk0} does not impose constraints on $m$ but fixes that the integration constant $\tau_B$ be zero, which is the FLRW solution \cite{plebanski2006introduction}.


\subsubsection{Mixed Power Law} 

The functional for this case reads as 
\begin{equation}
\label{MPLM}
f(T,B) = f_0 T^m B^n.
\end{equation}
As described in the previous case, we use \eqref{eq:antisymk0}, the ansatz \eqref{eq:Ydotansatz} and the substraction of Eqs.~\eqref{eq:eqrrk0},\eqref{eq:eqththk0}. The latter combined with Eq. \eqref{eq:Ydotansatz} give three possible constraints for the parameters $n,m$:  $n+m=1$, $(l-4)n-2m=0$, $n=2^{\frac{m}{2}}$. From this analysis, only $n+m=1$ provides a constant $l$ without yielding additional constraints on the radial function $Y$ or the mass function $M$. Therefore, we will consider this constraint for our study. Eq.~\eqref{eq:antisymk0} gives $l=1$ or $l=4$ and no additional constraints are imposed on $f_0$, but as we noticed in the previous section, the case $l=4$ implies no boundary term obtaining a vacuum solution. With these restrictions on the parameters we can derive the following equation for the mass--energy density using $l=1$, hence  Eq.~\eqref{eq:eqttk0} can be rewritten as
\begin{equation}\label{eq:rhomixed}
	8\pi \rho =-\frac{3^{m}2^n n f_0 M'}{Y' Y^2}.
\end{equation}
Notice that a positive mass-energy implies
\begin{equation}
	\frac{M'f_0}{Y'} \leq 0.
\end{equation}
Eq.~\eqref{eq:Ydotansatz} has the same solution as in GR
\begin{equation}
	Y=\left(\frac{9}{2}M(\tau-\tau_B(r))^2\right)^{\frac{1}{3}},
\end{equation} 
which in turn implies that the mass--energy density Eq.~\eqref{eq:rhomixed} has a continuous behaviour for all $r$. As described in the previous case, this solution does not modify the Hubble law, which is the case $l\neq 1$.

To extend the mixed power law model, we perform following mapping
\begin{equation}
	f(T,B)\mapsto \tilde f (T,B)= -T+ f(T,B),
\end{equation}
where we have a natural TEGR limit. As in the previous case from substracting Eqs.~ \eqref{eq:eqrrk0}, \eqref{eq:eqththk0}, here we obtain the same constraints for the parameters $n,m,$ namely: $n+m=1$, $(l-4)n-2m=0$, $n=2^{\frac{m}{2}}$. All three constraints provide a $l$ that does not impose additional restrictions on the radial function $Y$. Also, the constraints have as solution  $l=4$, and only $n+m=0$ has as an additional solution $l=1$. Therefore, this is the only case that deals with a non--zero boundary term. Under these arguments, Eq.~\eqref{eq:eqttk0} can be read as
\begin{equation}
	8\pi\rho=\frac{2M'}{Y^2Y'}-\frac{3^{m}2^n n f_0 M'}{Y' Y^2}.
\end{equation}
Notice that we have the usual GR term plus the term that we obtained from the mixed power law without the mapping described above. The mass term is
\begin{equation}
	M=4\pi \int \rho \left(1- \frac{3^{-m}2^{n-1} }{n f_0}\right)Y^2 Y'dr, 
\end{equation}
which produces a mass defect compared to the GR mass whenever $f_0 >0$, just as in $f(R)$ theories \cite{sussman2017lemaitre}. Positive definiteness must be analyzed case by case. 


\subsection{Accelerated cosmic expansion}
\label{sec:accexp}

As shown in Sec.~\ref{sec:struform} the proposed ansatzs do not provide a cosmic acceleration, therefore, a different ansatz can be introduced in order to mimic the LTB geometry plus a cosmological constant as
\begin{equation}\label{eq:Ydotanscosacc}
	\dot Y^2=\frac{2M}{Y^l}+LY^p,
\end{equation}
where $l,p$ are constants to be determined and $L$ is a free parameter that depends on time $L=L(\tau)$. This ansatz is inspired on the GR solution with a cosmological constant, in such case the evolution equation for the radial function $Y$ reads 
\begin{equation}\label{eq:LTBLam}
	\dot Y^2=\frac{2M}{Y}+\frac{\Lambda}{3}Y^2.
\end{equation}
In this case we recover the GR scenario with our proposed ansatz under $l=1, p=2$ and a constant $L$. Since these parameters are different than those considered in Sec.~\ref{sec:struform} we should select a particular set of parameters to analyse the consistency of such ansatz within a LTB geometry. Nevertheless, we can still analyse the behaviour of torsion and boundary terms, which in the particular case $l=1$ read
{\small
\begin{align}
	T&= \frac{4M'}{Y'Y^2}+\frac{2(p+1)LY^{p+\frac32}}{Y^{\frac{7}{2}}},\\
	B&= \frac{L}{(LY^{p+1}+2M)^2}\left((p+4)(p+1)Y^{3p}L^2+4(p+4)(p+1)M\left(MY^{p-2}+LY^{2p-1}\right)\right.\nonumber\\
	&\left.+\frac{6M'(4MY^{p-1}+4M^2Y^{-2}+LY^{2p})}{Y'}\right)+\frac{\dot L}{(LY^{p+1}+2M)^\frac52}\left(\left(\frac12(p+4)L^2 Y^{3p}\right.\right.\nonumber\\
	&\left.\left.-\frac{LM'Y^{2p}}{Y'}\right)Y^{\frac32}+\left(3(p+3)LMY^{2p}-\frac{2MM'Y^p}{Y'}\right)Y^\frac12 +4\left(p+\frac52\right)M^2Y^{p-\frac12}\right).
\end{align}}
 According to the latest observational surveys, the cosmological constant, $\Lambda$, has a small value, in the particular case in which we would consider our free parameter $L$ to resemble a cosmological constant a current time, the $B$ term can be shown to be still highly dependant on this parameter. Nevertheless, as the value should be small and we have several other variables such as the mass and the radial function $Y$, the effects of this cosmological constant type of behaviour will only be shown at non-linear orders, which is consistent with the work in \cite{Escamilla-Rivera:2019ulu}.

\section{Conclusions}
\label{sec:conc}

In this paper we have studied the effects of torsion-based models, namely $f(T,B)$ gravity, on the spherically symmetric LTB dust models, which is a simple exact solution providing a dust description of cold dark matter. 

We begin by discussing the LTB solution from Schwarzschild solution in GR, which provides a background consistency check to test the solutions in $f(T,B)$ theories and allows
an ansatz for the resulting PDE's. As part of this discussion, we briefly introduce the principles of TG in order to analyse the most general tetrad that recovers the LTB metric. In such analysis, the $\text{SO}(3)$ symmetry is Weitzenböck gauge-compatible as a consequence of a proper choice of parameters.
Also, we showed that the $tr$ component of the antisymmetric part of the field equations does not vanish identically, and therefore the symmetric part of the field equations and the non-zero component of the antisymmetric part of the equations have to be solved simultaneously for particular choices of $f(T,B)$. 

As we are interested in viable cosmological scenarios, we reduced our analysis to the spatially flat case and considered a couple of $f(T,B)$ functionals, e.g. modified versions of Power Law (\ref{PLM}) and Mixed Power Law (\ref{MPLM}) models, which have proved to be successful in reproducing cosmological scenarios such as the late cosmic acceleration and solve the $H_0$ tension. 

The motivations for studying inhomogeneous models in TG lies in the fact that we can construct toy models to analyse structure formation and the accelerated cosmic expansion in local regimes. To study the structure formation scenario, we proposed the ansatz (\ref{eq:Ydotansatz}) which is inspired by the standard Schwarzschild solution. In this case, we obtained that $l=4$ is a solution that recovers $f(T)$ gravity with an exact closed form of the radial function $Y$ and a continuous and regular equation for the mass-energy density in terms of the radial function $Y$ and the free parameters of the $f(T,B)$ functional. Nevertheless, this solution modifies the Friedmann equation and sets specific constraints to reproduce a viable structure formation according to the observations. In particular, the case $l \neq 4$ can be solved for $\alpha=0$, where the solution is FLRW with $l=1$. For the Mixed Power Law model, we consider $n+m=1$, $(l-4)n - 2m = 0$ and $n = 2^{\frac{m}{2}}$, where only the first option provides a constant $l$ without imposing additional constraints on $Y$ or $M$. 

Furthermore, we found two possibles solutions for $l=1$ and $l=4$, nevertheless, the case $l=4$ implies a vanishing boundary term and by consequence a vacuum solution. The case $l=1$ does not modify Hubble's law and gives the standard GR solution for the radial function $Y$. It is remarkable to mention that all constraints contain $l=4$ as a solution and only $n+m=1$ includes the case $l=1$. Here, the mass-energy density takes the form of the mass-energy density of GR plus an extra term. Here, this extra term produces a mass defect in comparison to GR when $f_0>0$. As an extension of this analysis, each of our set of parameters can be constrained using current observations, study that will be reported elsewhere. 


\section*{Acknowledgments}

CE-R acknowledges the Royal Astronomical Society as FRAS 10147. CE-R, A.A and  GAR-F are supported by PAPIIT-DGAPA UNAM Projects IA100220 and TA100122.
SN, A.A and GAR-F acknowledge financial support from SEP–CONACYT postgraduate grants program. RAS acknowledges support from PAPIIT-DGAPA RR107015. 
This work is part of the Cosmostatistics National Group (\href{https://www.nucleares.unam.mx/CosmoNag/index.html}{CosmoNag}) project.
The Authors thank J. Levi Said, S. Bahamonde and A. Golovnev  for enlightening discussions.

\bibliographystyle{unsrt}
\bibliography{references}

\end{document}